\documentclass[aps,amsfonts,showpacs]{revtex4-1}

\usepackage{graphicx}
\usepackage{natbib}
\usepackage{psfrag}
\usepackage{amssymb}
\usepackage{amsmath}
\usepackage{color}
\usepackage{hyperref}

\begin{document}

\def\be{\begin{equation}}
\def\ee{\end{equation}}
\def\bea{\begin{eqnarray}}
\def\eea{\end{eqnarray}}
\def\bml{\begin{mathletters}}
\def\eml{\end{mathletters}}
\def\b{\bullet}
\def\eqn#1{(~\ref{eq:#1}~)}
\def\no{\nonumber}
\def\av#1{{\langle  #1 \rangle}}
\def\m{{\rm{min}}}
\def\M{{\rm{max}}}
\newcommand{\ds}{\displaystyle}
\newcommand{\tc}{\textcolor}


\title{Critical dynamics of the jamming transition in one-dimensional nonequilibrium lattice-gas
models}
\author{Priyanka and Kavita Jain}
\affiliation{Theoretical Sciences Unit, Jawaharlal Nehru Centre for Advanced Scientific Research,
Jakkur P.O., Bangalore 560064, India}
\widetext
\date{\today}

\begin{abstract}
We consider several one-dimensional driven lattice-gas models that show a phase transition in the stationary state between a high-density fluid phase in which the typical length of a hole cluster is of order unity and a low-density jammed phase where a hole cluster of macroscopic length forms in front of a particle. Using a hydrodynamic equation for an interface growth model obtained from the driven lattice-gas models of interest here, we find  
that in the fluid phase, the roughness exponent and the dynamic exponent that, respectively, characterize the scaling of the 
saturation width and the relaxation time of the interface with the system size are given by the Kardar-Parisi-Zhang exponents. However, at the critical point, we show analytically that 
when the equal time density-density correlation function decays slower than inverse distance, the roughness exponent varies continuously  with a parameter in the hop rates but it is one half otherwise. Using these results and numerical simulations for the density-density autocorrelation function, we further find that the dynamic exponent $z=3/2$ in all cases.  
\end{abstract}

\maketitle

\section{Introduction}

Jamming is a ubiquitous phenomenon that disrupts the flow of vehicles 
on a city road \cite{Chowdhury:2000}, molecular motors in a biological cell  \cite{Leduc:2012}, fluids in narrow pipes and food grains through a hopper \cite{Liu:1998}. Usually one expects a jam to form at high densities and dissolve at lower densities; however, a jam can form at low particle densities also \cite{Loan:1998}. For example, on a bus route, if one of the buses gets delayed, the number of passengers waiting for it increases and so does the time interval during   which the bus halts. As a result, the bus gets delayed even more leading to a cluster of buses behind it 
even when few buses are serving on the affected bus route \cite{Loan:1998}. 
A similar situation occurs on an ant trail when the pheromone that guides the ant movement gets evaporated \cite{Chowdhury:2002}. Large headways in front of a jam  have been observed in experiments on ant trails \cite{John:2009} and camphor boats \cite{Heisler:2012} where the agents (ants and boats) form a platoon at low densities due to the presence of an external field, namely, pheromone and camphor concentration respectively.
Of course, a traffic jam can also form behind a slow car on a single-lane highway where overtaking is prohibited even when there are few vehicles \cite{Krug:1996,Evans:1996,Jain:2003}.

Here we study the stationary state dynamics of several driven lattice-gas models in one dimension that exhibit a jamming transition in the sense that the gap between the particles is of order unity when the total particle density is high. However, a long  headway forms in front of a particle while other particles get clustered behind it, when the density is lowered. A special case of the models studied here is the totally asymmetric simple exclusion process (TASEP) in which a particle hops to the left neighbor with rate one if it is empty. It is known that the jamming transition does not occur in this model and its steady state dynamics can be understood using its relationship to the Kardar-Parisi-Zhang equation for interface growth; specifically, the density-density autocorrelation function decays with time as $t^{-2/3}$  \cite{Kriecherbauer:2010,Halpin-Healy:2015}. The 
models of interest here are also driven and the particle interactions are hard core, but the hop rates in these models are such that a jamming 
transition can occur. Moreover, in the fluid phase and at the critical point, the hole clusters are not macroscopically long as in the TASEP.  Then it is natural to ask if the dynamical properties in the fluid phase and especially at the critical point 
differ from those in the TASEP.

The class of models that we consider here are related to a zero range process whose stationary state is known exactly \cite{Andjel:1982,Evans:2005} or a misanthrope process for which the exact steady state measure can be found in certain cases \cite{Thivent:1985,Evans:2014}. In these models, a jamming transition appears as a condensation transition where a macroscopically large mass cluster forms on a single site at high  densities but at low densities, the mass is distributed uniformly. The coarsening dynamics in the condensate phase have been studied for these models quite extensively  \cite{Krug:1996,Grosskinsky:2003,Godreche:2003,Jain:2003,Gupta:2007,Evans:2014}, but the
stationary state dynamics are largely unexplored (however, see \cite{Jain:2003,Godreche:2005}). It should be noted that in this picture, the questions relate to the properties of hole {\it clusters}, whereas here we are concerned with the holes themselves.

The relation to the zero range process or misanthrope process allows us to find the static density-density correlation function analytically. Recently, we studied the equal time density-density correlation function in the stationary state \cite{Priyanka:2014}, and it was shown that at the critical point, it decays with distance as a power law with a continuously varying exponent. Using this result and numerical simulations, we find that at the critical point, the autocorrelation function decays as a power law in time with an exponent that changes continuously with a parameter in the model when the static correlation function decays slower than inverse distance, but the TASEP behavior for the autocorrelation function, {\it viz.}, $t^{-2/3}$ decay holds otherwise. 

In the following section, we describe our basic model and review the known results. We then discuss the results for the static and dynamic correlation function in Secs.~\ref{equal} and \ref{unequal} respectively. Section \ref{general} is devoted to a discussion of two other models that exhibit jamming transition. We conclude with a summary of results and open questions in Sec.~\ref{concl}. 


\section{Model}
\label{model}

\subsection{Jamming transition}
\label{sub_model1}

\begin{figure}
\begin{center}
\includegraphics[width=1\textwidth]{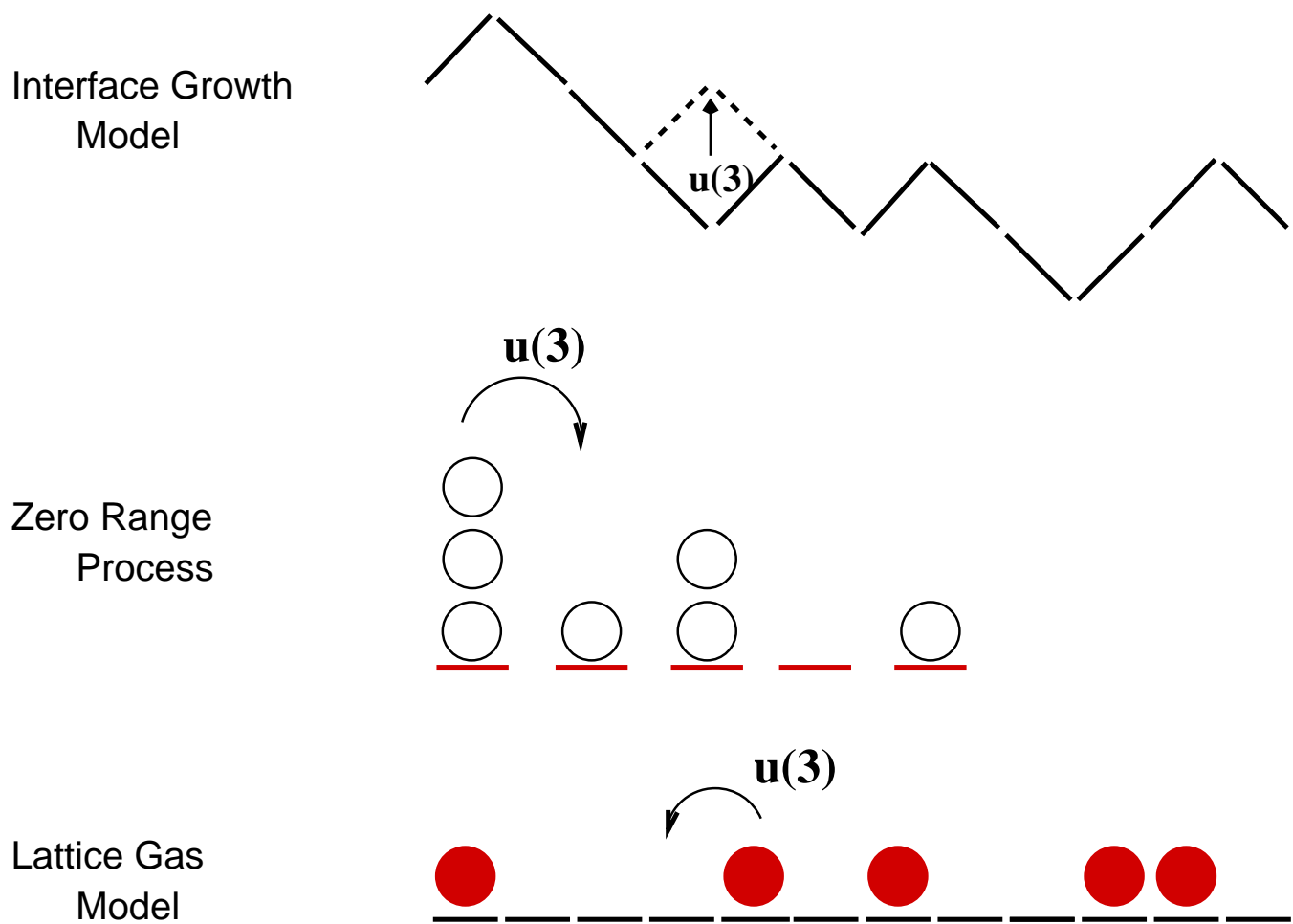}
\end{center}
\caption{Figure to illustrate the mapping between the lattice gas model (bottom), zero range process (middle) and interface growth model (top).}
\label{fig_mapping}
\end{figure}

We consider a system of hard core particles on a ring with $L$ sites in which the total number of particles $N$ is conserved. 
A particle attempts to hop to the empty site on the left with a rate $u(n)$, where $n > 0$ are the number of vacancies in front of it, see Fig.~\ref{fig_mapping} for an illustration. Let $\eta_i(t)$; denote the occupancy variable at site $i$ and time $t$ which takes the value $1$ if the site $i$ is occupied and zero otherwise. In continuous time, one can then write
\be
\frac{d \langle \eta_i \rangle }{dt}= \langle \sum_{m=0}^\infty \eta_{i+1} \eta_{i-(m+1)} u(m+1)\prod_{k=0}^{m}(1-\eta_{i-k}) - \sum_{m=1}^\infty \eta_{i} \eta_{i-(m+1)} u(m)\prod_{k=1}^{m}(1-\eta_{i-k}) \rangle ~, 
\ee
where the angular brackets denote averaging over initial conditions and independent stochastic trajectories.

Here we will consider the hop rate given by \cite{Evans:2005}
\be
u(n)=1+\frac{b}{n}~,~n > 0 ~,~
\label{hoprate}
\ee
where the parameter $b \geq 0$. 
When $b=0$, we obtain the totally asymmetric simple exclusion process (TASEP) \cite{Evans:2005,Kriecherbauer:2010}. For nonzero $b$, the rate $u(n)$ decreases with the increasing number of vacancies and therefore a particle with many holes in front of it hops at a slower rate and can cause a jam behind it. 
In fact, it has been shown that for $b > 2$, a phase transition occurs between a homogeneous phase in which the length of the typical hole cluster is of order unity and a jammed phase with a macroscopically large hole cluster, as the total particle density is decreased. 

To see this result, it is useful to consider a zero range process (ZRP) in which there is no restriction on the
number of particles that a site can support \cite{Evans:2005}. As shown in  Fig.~\ref{fig_mapping}, regarding a particle as a site and vacancies as mass clusters, the exclusion process described above with $L$ sites and $N$ particles maps to a ZRP with ${\cal L}=N$ sites and ${\cal M}=L-N$ particles (each of unit mass) in which a particle hops out to its right neighbor at a rate that depends on the number of particles at the departure site \cite{Evans:2005}. The density $\varrho={\cal M}/{\cal L}$ in the ZRP is related to the density $\rho=N/L$ in the exclusion process as
\be
\varrho=\frac{1}{\rho}-1 ~.
\ee
The exact mass distribution in the ZRP is known to be \cite{Evans:2005}
\be
{\cal P}(m_1,...,m_{\cal L})=Z_{{\cal L},{\cal M}}^{-1} ~f(m_1) ...  f(m_{\cal L}) ~,~
\label{products}
\ee
where the canonical partition function 
$Z_{{\cal L},{\cal M}}= \sum_{\{ m_j\}} f(m_1) ...  f(m_{\cal L}) ~\delta_{\sum_{j=1}^{\cal L} m_j,{\cal M}} $ and the weight factor 
\be
f(m)=\prod_{i=1}^m \frac{1}{u(i)}=\frac{f(m-1)}{u(m)} ~,~m \geq 1 ~,
\label{fmzrp}
 \ee
with $f(0)=1$. 
The grandcanonical partition function ${\cal Z}_{\cal L}(\omega)=\sum_{m=0}^\infty \omega^m Z_{{\cal L},m}= g^{\cal L}(\omega)$ where $g(\omega)=\sum_{m=0}^\infty \omega^m f(m)$.  Therefore the  marginal distribution ${\cal P}(m_1,...,m_s)$ that $s$ consecutive sites in the ZRP contain mass $m=\sum_{i=1}^s m_i$ is given by
\bea
{\cal P}(m_1,...,m_s) &=& \sum_{m_{s+1},...,m_{\cal L}} {\cal P}(m_1,...,m_{\cal L}) ~\delta_{\sum_{j=s+1}^{\cal L} m_j,{\cal M}-m} \\
&=& f(m_1) ... f(m_s) ~\frac{Z_{{\cal L}-s,{\cal M}-m}}{Z_{{\cal L},{\cal M}}} ~,
\eea
which, in the grandcanonical ensemble, reduces to
\bea
{\cal P}(m_1,...,m_s) &=& \prod_{i=1}^s \omega^{m_i} f(m_i) ~\frac{{\cal Z}_{{\cal L}-s}}{{\cal Z}_{{\cal L}}}  \\
&=& \prod_{i=1}^s \frac{\omega^{m_i} f(m_i)}{g(\omega)} ~.
\label{pmdefn}
\eea
In particular, the single site mass distribution is given by 
\be
p(m)=\frac{\omega^{m} f(m)}{g(\omega)} ~.
\label{pm1}
\ee
In the above equations, the fugacity $\omega$ is determined by the mass conservation equation,
\be
\varrho=\frac{\omega}{{\cal L}} \frac{\partial \ln {\cal Z}_{\cal L}}{\partial \omega}=\omega \frac{\partial \ln g(\omega)}{\partial \omega} ~.
\label{conser}
\ee
From (\ref{pm1}) and (\ref{conser}), we also have 
\be
\sigma^2= \omega \frac{\partial \varrho}{\partial \omega} = - \frac{\omega}{\rho^2} ~\frac{\partial \rho}{\partial \omega}~,
\label{variance}
\ee
where, $\sigma^2= \langle m^2 \rangle - \langle m \rangle^2$ is the mass variance in ZRP. The above equation shows that as the particle density $\rho$ is decreased, the fugacity increases. 

For the hop rate (\ref{hoprate}), the generating function $g(\omega)={_2}F_1(1,1;b+1;\omega)$ is the Gauss hypergeometric function which  converges for $ |\omega| < 1$ \cite{Abramowitz:1964}. Therefore, if a solution for the fugacity $0 < \omega < 1$ exists for all $0 < \rho < 1$, the stationary state is in the fluid phase since  the distribution of the hole cluster length is an exponential, see (\ref{pm1}). However, when $b > 2$, the maximum value $\omega=1$ is reached at a nonzero critical density $\rho_c=(b-2)/(b-1)$ as can be seen using (\ref{conser}) and that the weight $f(m) \sim m^{-b}$ for large $m$. 
When the particle density is lowered below $\rho_c$, since (\ref{conser}) yields the density as $\rho_c$, the excess density is compensated by a hole cluster of finite density. 
Thus for $b > 2$, the system is in a homogeneous phase for $\rho > \rho_c$, while for $\rho < \rho_c$, the homogeneous phase at particle density $\rho_c$ 
coexists with a macroscopically long hole cluster of length {$L (\rho_c-\rho)/\rho_c$}.  

The exclusion process described at the beginning of this section can also be  related to an interface growth model. 
As shown in  Fig.~\ref{fig_mapping}, an interface height profile can be obtained by associating an upward (downward) slope with a particle (hole) in the exclusion process. The interface evolves in time by flipping a local valley to a hill with a rate that depends on the number of upward slopes on the left. It is convenient to work with a height profile with zero average slope, and we therefore define 
the height $h_j(t)$ of the interface at the substrate site $j$ and time $t$ as 
\be
h_j(t)= \sum_{i=1}^j {\eta_i}(t) -\rho j ~.
\label{hnmap}
\ee


\subsection{Density-density correlation function}
\label{sub_model2}

Consider the unequal time density-density correlation function defined as 
\bea
S(r,t) = L^{-1} \sum_{i=1}^L \left[ \langle \eta_i(0) \eta_{i+r}(t) \rangle -\rho^2  \right]~,
\label{defnS} 
\eea
where the angular brackets denote an average over initial conditions that are chosen from the {\it stationary state ensemble} and independent stochastic histories. We will study the equal time correlation function $S(r,0)$ in Sec.~\ref{equal} and the autocorrelation function  $S(0,t)$ in Sec.~\ref{unequal}. 
The correlation function $S(r,t)$ is related to the height-height correlation function defined as \cite{Kriecherbauer:2010,Halpin-Healy:2015}
\be
C(r,t)=  L^{-1} \sum_{i=1}^L \left[ \langle (h_{i+r}(t)-h_i(0))^2 \rangle \right]~,
\label{defnC}
\ee
via the relation \cite{Prahofer:2004}
\bea
S(r,t) = \frac{1}{2} \frac{\partial^2 C(r,t)}{\partial r^2} ~,
\label{relnSC}
\eea
which can be obtained on using (\ref{hnmap}) in the derivative of $C(r,t)$ with respect to $r$. 

The correlation function $C(r,t)$ has been studied for several interface growth models, and found to grow with time as $t^{2 \beta}$ initially and 
saturate to $r^{2 \alpha}$ over a time scale that grows as $r^z$. In other words, it is of the scaling form \cite{Prahofer:2004} 
\be
C(r,t)= t^{2 \beta} ~{\cal C} \left( \frac{r}{t^{1/z}}\right) ~,
\ee
where the scaling function ${\cal C}(x)$ is a constant when $x \gg 1$ and $x^{2 \alpha}$ for $x \ll 1$, and  $\alpha=\beta z$. 
Correspondingly, for the density-density correlation function, we have
\be
S(r,t)=t^{-\frac{2 (1-\alpha)}{z}} ~{\cal S} \left( \frac{t}{r^z} \right) ~,
\label{Srtscaling}
\ee
where the scaling function ${\cal S}(x)$ is a constant for $x \ll 1$ and decays to zero for $x \gg 1$. 

The height model corresponding to the TASEP ($b=0$) is known to obey the Kardar-Parisi-Zhang (KPZ) equation \cite{Kardar:1986} for which it is known that the dynamic exponent $z=3/2$, the roughness exponent $\alpha=2-z=1/2$ and the growth exponent $\beta=1/3$ (see also Appendix~\ref{app_hydro}), and therefore $S(0,t) \propto t^{-2/3}$ for large systems as indeed verified in recent numerical simulations \cite{Daquila:2011}. Our objective here is to study the exclusion process with hole-dependent hop rates at the critical point and determine if the dynamics differ from that in the $b=0$ case. 
The autocorrelation function is studied by simulating the driven lattice gas model via sequential updating. In most cases, we worked with large systems of size $L=10^{4}$ or larger, and the data were averaged over more than $10^4$ independent initial conditions in the stationary state. 
Our results for $S(r,0)$ and $S(0,t)$ are described in the following two sections. Although in most of the article, we focus on the exclusion process described at the beginning of this section,  
we also discuss two other models that exhibit a jamming transition in Sec.~\ref{general}. 

\section{Equal time correlation function}
\label{equal}

\subsection{Density-density correlation function}

In this subsection, we are interested in the
static two-point correlation function $S(r,0)$. 
In  a recent work \cite{Priyanka:2014}, this correlation function was calculated  
in the fluid phase and at the critical point using an exact expression for it.  It was shown that at the critical point, the equal time correlation  function decays as a power law with a continuously varying exponent, see (\ref{Sr0cp}) below. Here we give a simpler derivation of this result.

Consider the probability $P_s(r-s)$ that $s$ consecutive sites in the ZRP have $r-s$ particles. It is then easy to see that in the exclusion process, the probability that two sites at a distance $r$ are occupied is given by 
\bea
\langle \eta_i(0) \eta_{i+r} (0)\rangle &=& \rho \sum_{s=1}^r P_{s} (r-s) ~.
\label{homocorr}
\eea
The factor $\rho$ on the RHS of the above equation appears because the site $i$ (in the
exclusion process) must be occupied. The generating function of the 
correlation function defined as $G(y)=\sum_{r=0}^{\infty}y^r S(r,0)$ is then given by 
\bea
G(y)&=&\rho \sum_{r=0}^{\infty}y^r  \sum_{s=1}^r P_{s} (r-s)-\sum_{r=0}^{\infty}y^r\rho^2 \\
&=& \rho \sum_{r=0}^\infty y^r \sum_{m=0}^\infty y^m P_r(m)-\sum_{r=0}^{\infty}y^r\rho^2 ~.
\label{Gy1}
\eea
But due to (\ref{pmdefn}), we have 
\bea
 \sum_{m=0}^\infty y^m P_r(m) &=&  \sum_{m=0}^\infty y^m \sum_{m_1,...,m_r} {\cal P}(m_1,...,m_r) ~\delta_{\sum_{j=1}^r m_j,m}\\
&=& \left(\frac{g(y\omega)}{g(\omega)}\right)^r ~.
\label{sumSr0}
\eea
Using the last expression in (\ref{Gy1}), we immediately obtain the result (32) in \cite{Priyanka:2014} for the generating function of the correlation function:
\be
G(y)= \frac{\rho}{1-y~\frac{g(y\omega)}{g(\omega)}} - \frac{\rho^2}{1-y} ~.
 \label{Prob_gen_Corr}
 \ee
The large distance behavior of the correlation function can be found by expanding $G(y)$ about $y=1$, and at the critical point, we have \cite{Priyanka:2014}
\be
 S(r,0)= \frac{\rho_c^2 \Gamma(b-1)}{r^{b-2}} ~,~ b > 2 ~.
\label{Sr0cp}
\ee
\subsection{Height-height correlation function and roughness exponent}

We now use the result obtained in the preceding subsection to find the equal time height-height correlation function $C(r,0)$ defined in (\ref{defnC}) which grows as $r^{2 \alpha}$ \cite{Kriecherbauer:2010,Halpin-Healy:2015}. 
Using (\ref{hnmap}) in (\ref{defnC}), we have 
\be
C(r,0)=\rho (1-\rho) r+ 2 \sum_{k=1}^r (r-k) S(k,0) ~.
\label{CSreln}
\ee
For $b=0$ (TASEP), as $S(k,0)$ vanishes for all $k > 0$ in the thermodynamic limit, the above correlation function increases linearly with $r$ and therefore the roughness exponent $\alpha$ equals $1/2$  \cite{Kriecherbauer:2010,Halpin-Healy:2015}. For $b > 2$, at the critical point, we get 
\be
C(r,0)=\rho_c (1-\rho_c) r+ {2 \rho_c^2 \Gamma(b-1)} ~\left(r H_r^{(b-2)}-H_r^{(b-3)} \right) ~,
\label{Cr0}
\ee
where we have used (\ref{Sr0cp}). In the above equation, 
$H_n^{(x)}=\sum_{i=1}^n i^{-x}$ is a harmonic number of order $x$ and whose large $n$ behavior is given by \cite{Abramowitz:1964}
\be
H_n^{(x)}=\frac{n^{1-x}}{1-x}+\zeta(x)+{\cal O}(n^{-x}),
\ee
where $\zeta(x)=\sum_{k=1}^\infty k^{-x}$ is the Riemann zeta function. Using the above equation in (\ref{Cr0}), we find the  leading and two subleading terms in the height-height correlation function to be 
\be
\label{Cr0multi}
{C(r,0)=} 
\begin{cases}
c_1 r^{4-b}+ c_2 r +c_3 &,~ b \neq 3,4\\
\left[2 \rho_c^2 \Gamma(b-1) \right] r \ln r+  c_2' r       &,~ b=3 \\
c_2 r -2 \rho_c^2 \Gamma(b-1) \ln r &,~ b=4 ~,
\end{cases}
\ee
where the coefficients $c_1=2 \rho_c^2 \Gamma(b-1) ((3-b) (4-b))^{-1}, c_2=\rho_c (1-\rho_c)+2 \rho_c^2 \Gamma(b-1) \zeta(b-2), c_2'=\rho_c (1-\rho_c)+2 \rho_c^2 \Gamma(b-1) (\gamma_{EM}-1), c_3=-2  \rho_c^2 \Gamma(b-1) \zeta(b-3)$ and  $\gamma_{EM} \approx 0.577$ is the Euler-Mascheroni constant. It can be verified that the correlation function $S(r,0)$ in (\ref{Sr0cp}) is obtained using the above result in (\ref{relnSC}). 

Equation (\ref{Cr0multi}) shows that for large $r$, the height variance grows  as $r^{4-b}$ for $2 < b < 3$ and linearly for $b > 3$. Thus the interface is rougher in the former case, and the growth of the interface width with distance can be characterised by the roughness exponent given by 
\be
{\alpha =} 
\begin{cases}
(4-b)/2 &,~ 2 < b < 3 \\
1/2 &,~ b > 3 ~.
\end{cases}
\label{alphaexpo}
\ee
The subleading terms in the correlation function $C(r,0)$ are useful in understanding the dynamical behavior of the correlation function which we discuss in the next section. 

\section{Unequal time correlation function}
\label{unequal}

\begin{figure}
\begin{center}
\includegraphics[width=1\textwidth]{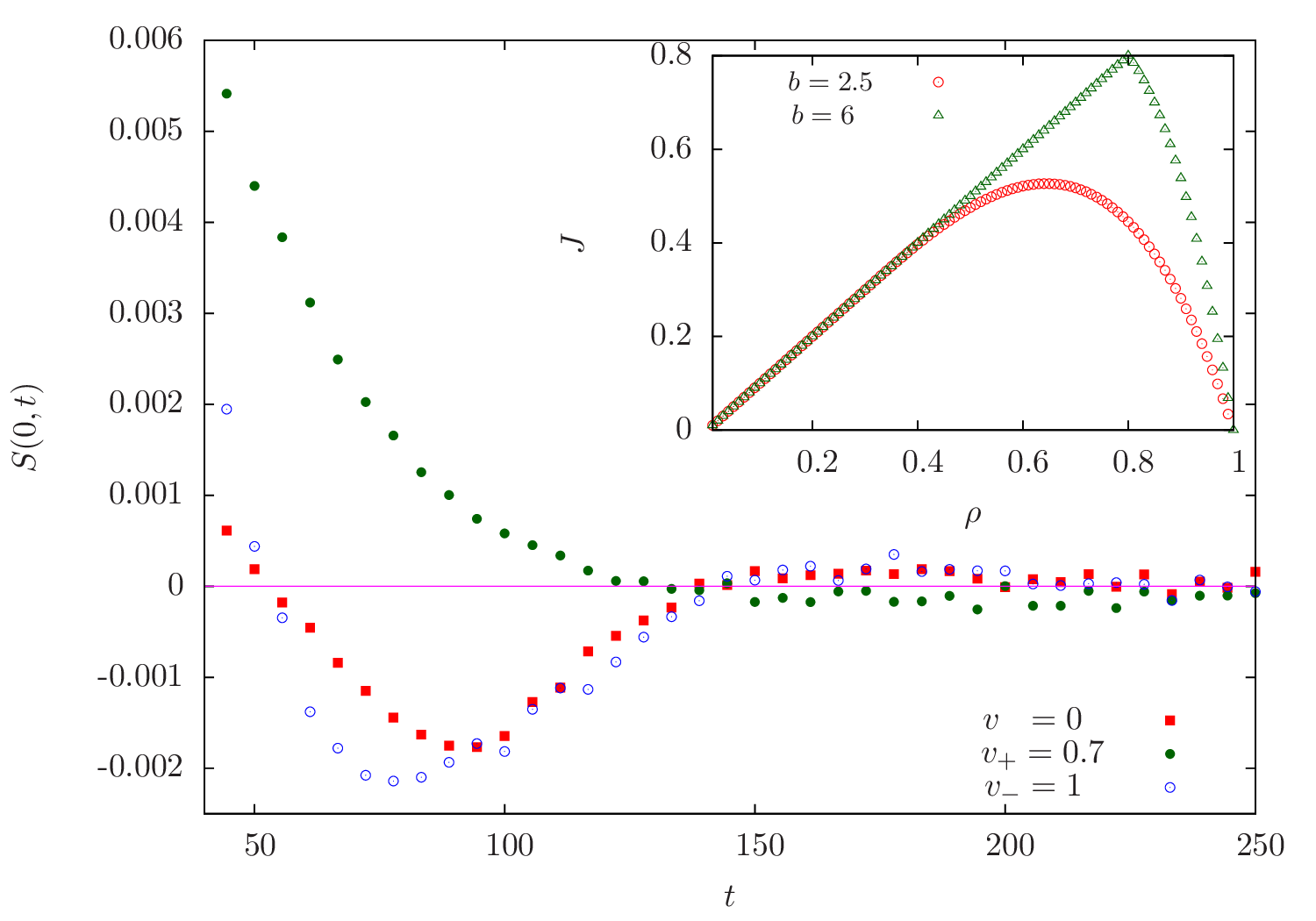}
\end{center}
\caption{Autocorrelation function $S(0,t)$ defined in (\ref{defnS2}) at the critical point for $b=7/2$ and $L=100$ to show that the density fluctuations move with the speed $v_+$ given by (\ref{kinspeed}). The inset shows the stationary state current (\ref{currentexpr}) as a function of the particle density for two values of the parameter $b$. The critical density $\rho_c=1/3$ and $4/5$ for $b=5/2$ and $6$ respectively.}
\label{fig_current}
\end{figure}

We now study the autocorrelation function 
at the critical point. The results of our numerical simulations are shown in the inset of Fig.~\ref{fig_current} for the autocorrelation function defined in (\ref{defnS}) and we find that it oscillates in time. 
This is because the density fluctuations move with a nonzero speed $v= \partial J/\partial \rho$ \cite{Lighthill:1955} (also, see Appendix~\ref{app_hydro}) where $J$ is the stationary state current. 
These oscillations (with time period $L/v$) can be eliminated by working in the rest frame of the density fluctuations \cite{Gupta:2007}, and we therefore consider 
\bea
S(r,t) = \langle \eta_i(0) \eta_{i+v t+r}(t) \rangle -\rho^2   ~.
\label{defnS2}
\eea

To find the speed $v$, we first calculate the stationary state current as follows. Since a particle hops to the left empty neighbor with a rate that depends on the number of vacancies in front of it, the stationary state current in the bond connecting the sites $i-1$ and $i$ is given by
\bea
J(\rho) &=& \langle u(1) \eta_{i-2} (1-\eta_{i-1}) \eta_{i} + u(2) \eta_{i-3} (1-\eta_{i-2}) (1-\eta_{i-1}) \eta_{i} + ...\rangle ~\\
&=& \langle\sum_{m=1}^{\infty}u(m)\eta_{i}\eta_{i-(m+1)}\prod_{k=1}^{m}(1-\eta_{i-k})\rangle.
\eea
On mapping the particle configurations appearing on the RHS of the above equation to the ZRP as described in Sec.~\ref{model}, we find that 
\bea
J(\rho)&=& \rho \sum_{m=1}^\infty u(m) p(m) \\
&=& \rho \sum_{m=1}^\infty u(m) \frac{\omega^m f(m)}{g(\omega)} \\
&=& \rho ~\omega(\rho)  ~,
\label{currentexpr}
\eea
since $p(m)$ is given by (\ref{pm1}) and $f(m)$ obeys (\ref{fmzrp}). 
As shown in the inset of Fig.~\ref{fig_current}, the current is a nonmonotonic function of density: it increases linearly in the jammed phase and reaches a maximum at $\rho \geq \rho_c$,  see Appendix~\ref{app_curr} for details. 

In the low density phase, since the fugacity equals one, the current is simply given by $\rho$ and thus the speed of the density fluctuations for $\rho < \rho_c$ is unity for all $b > 2$. In the high density phase where $\omega < 1$, using (\ref{variance}), we obtain 
\be
v_+= \omega - \frac{\omega}{\rho \sigma^2} ~.
\ee
At the critical point, since the mass distribution $p(m) \sim m^{-b}$, the variance $\sigma^2$ diverges when $2 < b < 3$ and remains finite otherwise. As a result, the speed is not a continuous function of density for $b > 3$ since
\begin{align}
{v_+(\rho_c) =} 
\begin{cases}
1 &~,~ 2 < b < 3 \\
1-\frac{(b-2) (b-3)}{b-1} &~,~ b > 3 ~, 
\end{cases}
\label{kinspeed}
\end{align}
whereas, as mentioned above, $v_-(\rho_c)=1$ for $b > 2$ \cite{Gupta:2007}. 
Thus, for $b > 3$, 
it is not clear whether one should use $v_+$ or $v_-$ in (\ref{defnS2}) at the critical point. Our numerical simulations show that the correlation function $S(0,t)$ does not oscillate in time when $v_+$ is used, see Fig.~\ref{fig_current}. To rationalize this observation, we note that the speed $v$ is determined by a steady state property (namely, the first derivative of the steady state current). But in the stationary state, a typical hole cluster is not macroscopically long (see  Sec.~\ref{sub_model1}) and therefore the properties of the critical state are similar to that in the fluid phase. In the following, we will set $v=v_+$ which is given by (\ref{kinspeed}). 

\subsection{When $b > 3$}

\begin{figure}
\begin{center}
\includegraphics[width=1\textwidth]{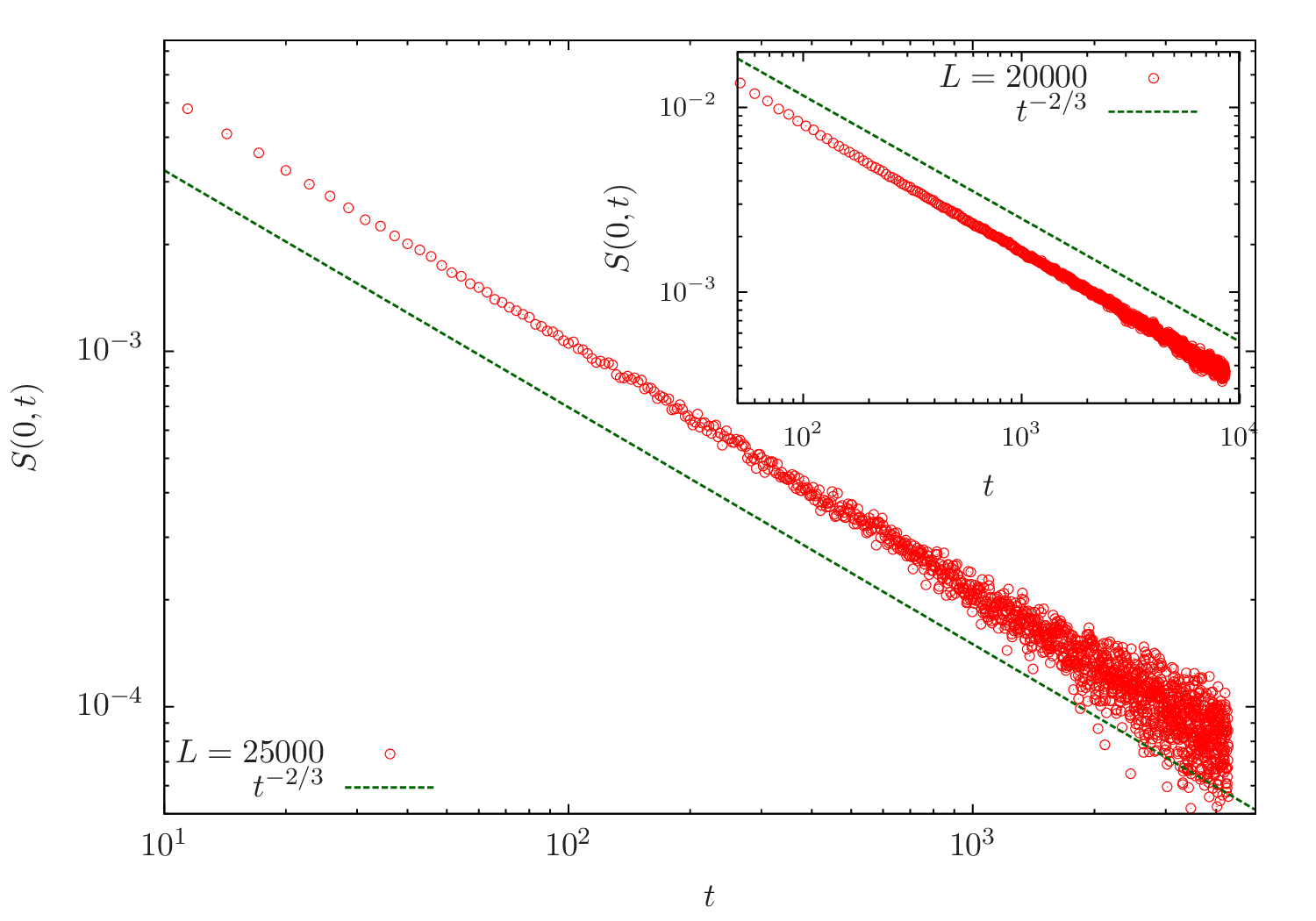}
\end{center}
\caption{Temporal decay of the autocorrelation function $S(0,t)$ defined in (\ref{defnS2}) at the critical point for the exclusion process with hop rate (\ref{hoprate}) and $b=6$. 
The inset shows the $t^{-2/3}$ decay of the autocorrelation function in the fluid phase for $b=5/2$ at $\rho=1/2$.}
\label{fig_b6}
\end{figure}

To get an insight into the dynamics, we write down an equation for the  height of the interface model (explained in Sec.~\ref{sub_model1}) in Appendix~\ref{app_hydro} using a standard prescription for $\rho \geq \rho_c$. The coefficients appearing in this equation are related to the second and third cumulants of the hole cluster distribution $p(m)$ in (\ref{pm1}). In the fluid phase where the distribution of the hole clusters is an exponential, we find that the KPZ equation is obtained for all $b > 2$ and therefore the autocorrelation function $S(0,t) \propto t^{-2/3}$, as shown in the inset of Fig.~\ref{fig_b6}. 

However, at the critical point where the hole cluster distribution $p(m) \sim m^{-b}$, we find that the interface height profile obeys the KPZ equation when $b > 4$. We thus expect that the density-density autocorrelation function $S(0,t)$ decays as $t^{-2/3}$ as in the standard KPZ case. Our numerical results shown in Fig.~\ref{fig_b6} for $b=6$ are indeed  consistent with this expectation.

\begin{figure}
\begin{center}
\includegraphics[width=1\textwidth]{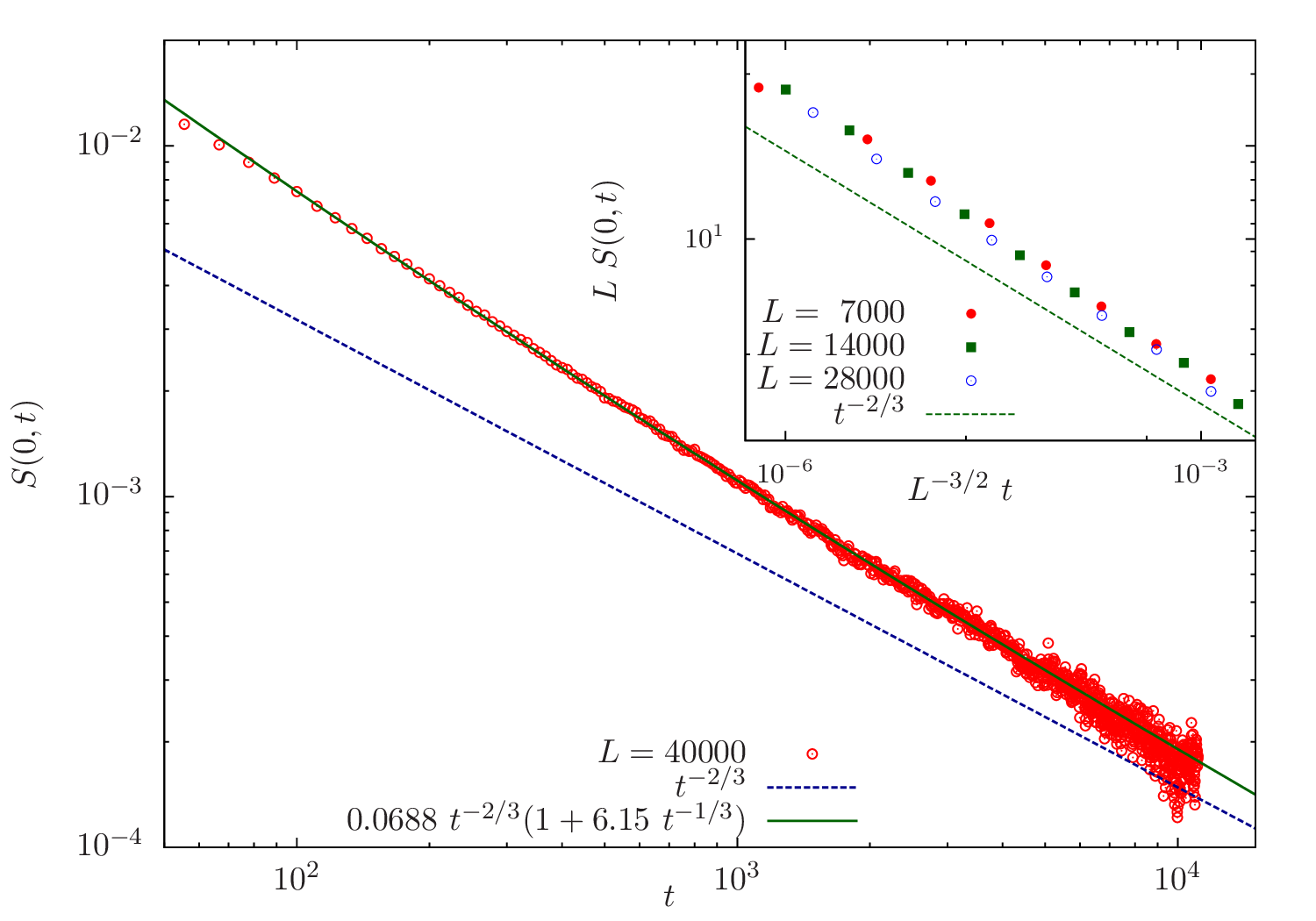}
\end{center}
\caption{Temporal decay of the autocorrelation function $S(0,t)$ defined in (\ref{defnS2}) at the critical point for the exclusion process with hop rate (\ref{hoprate}) and $b=7/2$. The inset shown the data collapse for various system sizes for $b=3.8$ at the critical density.}
\label{fig_b3p5}
\end{figure}

We next turn to the case when $3 < b < 4$. Although the roughness exponent in (\ref{alphaexpo}) is $1/2$ in this parameter regime, as the third cumulant of $p(m)$ diverges for $b < 4$, our hydrodynamic description in Appendix~\ref{app_hydro} breaks down. However, our numerical data for the autocorrelation function shown in Fig.~\ref{fig_b3p5} is consistent with the $t^{-2/3}$ decay at large times which means that the dynamic exponent $z=3/2$ here as well. This is also verified by writing (\ref{Srtscaling}) as $r S(r,t)= x^{-2/3} {\cal S}(x)~,~x=t\,r^{-3/2}$ and obtaining a data collapse for three system sizes as shown in the inset of Fig.~\ref{fig_b3p5}. 
The slow convergence of $S(0,t)$ to the asymptotic behavior indicates that the subleading corrections to the leading behavior may be strong. From (\ref{Cr0multi}) for the static correlation function, we see that the subleading correction grows as $r^{4-b}$ for $b > 3$. On using this result and the discussion in Sec.~\ref{sub_model2}, we find that  
\be
S(0,t) \propto \frac{1}{t^{\frac{2}{3}}} \left(1+ \frac{A}{t^{\frac{2 (b-3)}{3}}} \right) ~,
\ee
where $A$ is a constant. Thus for $b=7/2$, the correction to the leading order behavior is expected to decay slowly as $t^{-1/3}$; the numerical data fits well to the above equation as shown in Fig.~\ref{fig_b3p5}. 

\subsection{When $2 < b < 3$}

\begin{figure}
\begin{center}
\includegraphics[width=1\textwidth]{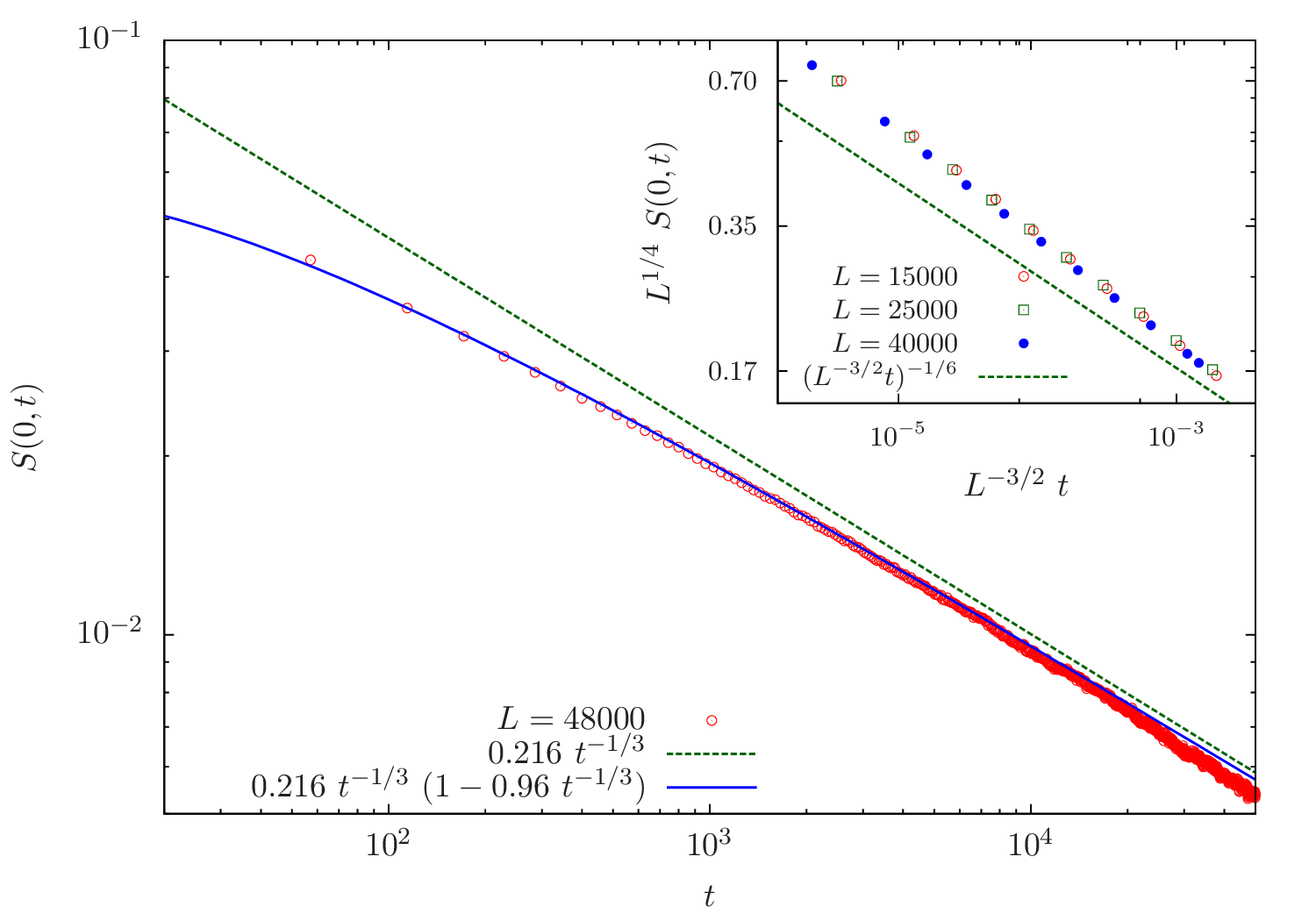}
\end{center}
\caption{Temporal decay of the autocorrelation function $S(0,t)$ defined in (\ref{defnS2}) at the critical point for the exclusion process with hop rate (\ref{hoprate}) and $b=5/2$. The inset shown the data collapse for various system sizes for $b=9/4$ at critical density.}
\label{fig_b2p5}
\end{figure}

As in the $3 < b < 4$ case, here our hydrodynamic description also breaks down, the reason for which can be traced to the slow decay of the hole cluster distribution. In addition, the static correlation function $C(r,0)$ increases superlinearly, see (\ref{Cr0multi}). As shown in the inset of Fig.~\ref{fig_b2p5}, we obtain a good data collapse for various system sizes using $\alpha=(4-b)/2$ and $z=3/2$ in (\ref{Srtscaling}). Thus, here the autocorrelation function decays with time as a power law with a continuously varying exponent given by $2(b-2)/3$.  Moreover, on taking the subleading corrections to the static correlation function into account, we obtain
\be
S(0,t) \propto \frac{1}{t^{\frac{2 (b-2)}{3}}} \left(1+ \frac{A'}{t^{\frac{2 (3-b)}{3}}} \right) ~,
\ee
which provides a good fit to the numerical data as shown in Fig.~\ref{fig_b2p5} for $b=5/2$. 


\section{Jamming transition and dynamics in other models}
\label{general}

The main conclusions so far 
are that at the critical point, in a certain parameter regime, the roughness exponent is different from that in the TASEP 
but the dynamic exponent $z=3/2$ holds throughout. 
Below we discuss two more models that show a similar jamming transition and for which 
the above results continue to hold. 


\subsection{Generalised exclusion process with hole-dependent rates}
\label{sub_misan}

We now discuss a class of driven lattice gas models in which a
particle hops to the empty left neighbor at a rate $u(m,m')$ where $m (m')$ 
denotes the number of vacancies on its left (right). This model is
obtained, via the mapping described in Sec.~\ref{model}, from a
misanthrope process \cite{Evans:2014} in which a site can be occupied
by any number of particles and a particle hops
out of a site with mass $m$ to its right neighbor at a rate $u(m,m')$ when there are $m'$ particles on the landing site.  
When the hop rate does not depend on the mass at the target site,  we obtain a ZRP as already discussed in Sec.~\ref{model}. 
For the misanthrope process, it has been shown that the mass distribution is of the product form  (\ref{products}) with $f(m)$ given by \cite{Thivent:1985,Evans:2014}
\be
f(m)= f(0) \left(\frac{f(1)}{f(0)} \right)^m \prod_{i=1}^m \frac{u(1,i-1)}{u(i,0)} ~,
\label{fmmisanth}
\ee
and (\ref{pmdefn})-(\ref{variance}) hold, provided certain conditions on the hop rates are satisfied.

Here we consider such a case with hop rates given by  \cite{Evans:2014}
\be
u(m,m')=(v(m)-v(0))~v(m') ~,
\label{ummmisa}
\ee
where 
\be
\frac{2}{3} <  v(0) < 1,~~~~~v(m)=1+\frac{1}{m+1}~,~m > 0  ~.
\label{vmmisa}
\ee
Thus a particle hops to the left empty site with a lower rate if there are large empty regions on either side. 
For the range of $v(0)$ mentioned above, a jamming transition similar to that described in Sec.~\ref{model} occurs as detailed in Appendix~\ref{app_misan}. 
Using (\ref{gzmis1}) for $g(\omega)$ in the expression (\ref{Prob_gen_Corr}) for the generating function of the static density-density correlation function and expanding about $y=1$, we find that  $G(y) \propto (1-y)^{c-6}$ at the critical point, where
\be
c=\frac{3-2 v(0)}{1-v(0)} > 5 ~, 
\ee
due to (\ref{vmmisa}). 
upon inverting the Laplace transform \cite{Priyanka:2014}, we finally obtain 
\be
S(r,0)=\frac{\varrho_c \rho_c^3 (\Gamma(c-2))^2}{4 \Gamma(c-3)} \times \frac{1}{r^{c-5}} ~,
  \label{misan_Cr}
  \ee
where the critical particle density $\rho_c$ and the related mass density  $\varrho_c$ are given in Appendix~\ref{app_misan}. On comparing this result with those in Sec.~\ref{equal}, we immediately see that the roughness exponent is equal to $(7-c)/2$ for $5 < c <6$ and $1/2$ for $c > 6$. 

Furthermore, assuming that the dynamic exponent $z=3/2$ here as well, we expect that the autocorrelation function behaves as 
\begin{align}
{S(0,t) \propto } 
\begin{cases}
{t^{-\frac{2 (c-5)}{3}}} \left(1+ {\alpha_1} ~{t^{-\frac{2 (6-c)}{3}}} \right)&~,~ 5 < c < 6 \\
{t^{-\frac{2}{3}}} \left(1+ {\alpha_2}~{t^{-\frac{2 (c-6)}{3}}} \right) &~,~ 6 < c < 7 \\
{t^{-\frac{2}{3}}} \left(1+ {\alpha_3}~{t^{-\frac{2}{3}}} \right)&~,~ c > 7 ~,
\end{cases}
\label{S0tmis}
\end{align}
where $\alpha_i$'s are constants. 
Our numerical results for the autocorrelation function shown in Fig.~\ref{fig_misant} for $c=5.5$ and $9$ in the rest frame of the density fluctuations (see Appendix~\ref{app_misan}) are in agreement with these expectations. 


\subsection{Exclusion process with particlewise disorder}
\label{sub_dis}

We again consider a system of hard core particles on a ring in which a particle hops to the empty left neighbor. However, the hop rate of a particle 
is now a random variable, and it is independent of the length of the  hole clusters on either side \cite{Krug:1996,Evans:1996,Jain:2003}. The hop rates are  chosen independently from a common distribution,  
\be
q(u) =\frac{1+\gamma}{(1-{\tilde u})^{1+\gamma}}(u-{\tilde u})^{\gamma}~,~\gamma > 0 ~,
\label{disdist}
\ee
 where $0 < {\tilde u} < u < 1$.  This model can be mapped to a ZRP with sitewise disorder in which the mass distribution ${\cal P}(m_1,...,m_{\cal L}) \propto \prod_{i=1}^{\cal L} f_i(m_i)$ where the (site-dependent) weight factor $f_i(m)=u_i^{-m}$. The grandcanonical partition function ${\cal Z}_{\cal L}= \prod_{i=1}^{\cal L} g_i(\omega)$ where $g_i(\omega)=u_i/(u_i-\omega)$. As before, the fugacity is determined by the particle conservation equation (\ref{conser}), 
\bea
\varrho = \frac{1}{{\cal L}} \sum_{i=1}^{\cal L} \frac{\omega}{u_i-\omega} ~.
\label{disconser}
\eea
Here, the fugacity is bounded above by the minimum hop rate and therefore, as before, with decreasing particle density, the fugacity decreases until it reaches ${\tilde u}$ (in the thermodynamic limit). Thus a jamming transition occurs 
 between a high-density homogeneous phase and a low-density jammed phase in which a macroscopically large hole cluster forms in front of the particle with minimum hop rate, viz., ${\tilde u}$ at $\rho_c=\gamma \frac{1-{\tilde u}}{{\tilde u}+\gamma}$ \cite{Krug:1996,Evans:1996}.

As explained in Appendix~\ref{app_dis}, at the critical point, the generating function of the equal time density-density correlation function after averaging over the hop rate distribution works out to be 
\be
\overline{G(y)}=\frac{\rho_c}{1-\overline{\frac{y~g({\tilde u}
      y)}{g({\tilde u})}}}-\frac{\rho_c^2}{1-y}  ~,
\label{disGy}
 \ee
where overbar indicates the average w.r.t. the distribution $q(u)$. The correlation function then turns out to be 
 \be
\overline{S(r,0)} =\rho_{c}^{3}(1+\gamma)\left(\frac{{\tilde u}}{1-{\tilde u}}\right)^{1+\gamma}\frac{\Gamma(\gamma)}{r^{\gamma}} ~,
 \ee
Using this expression in (\ref{CSreln}), we find that the roughness exponent is given by $(2-\gamma)/2$ when  $0 < \gamma < 1$ but is $1/2$ for $\gamma > 1$. 

\begin{figure}
\begin{center}
\includegraphics[width=1\textwidth]{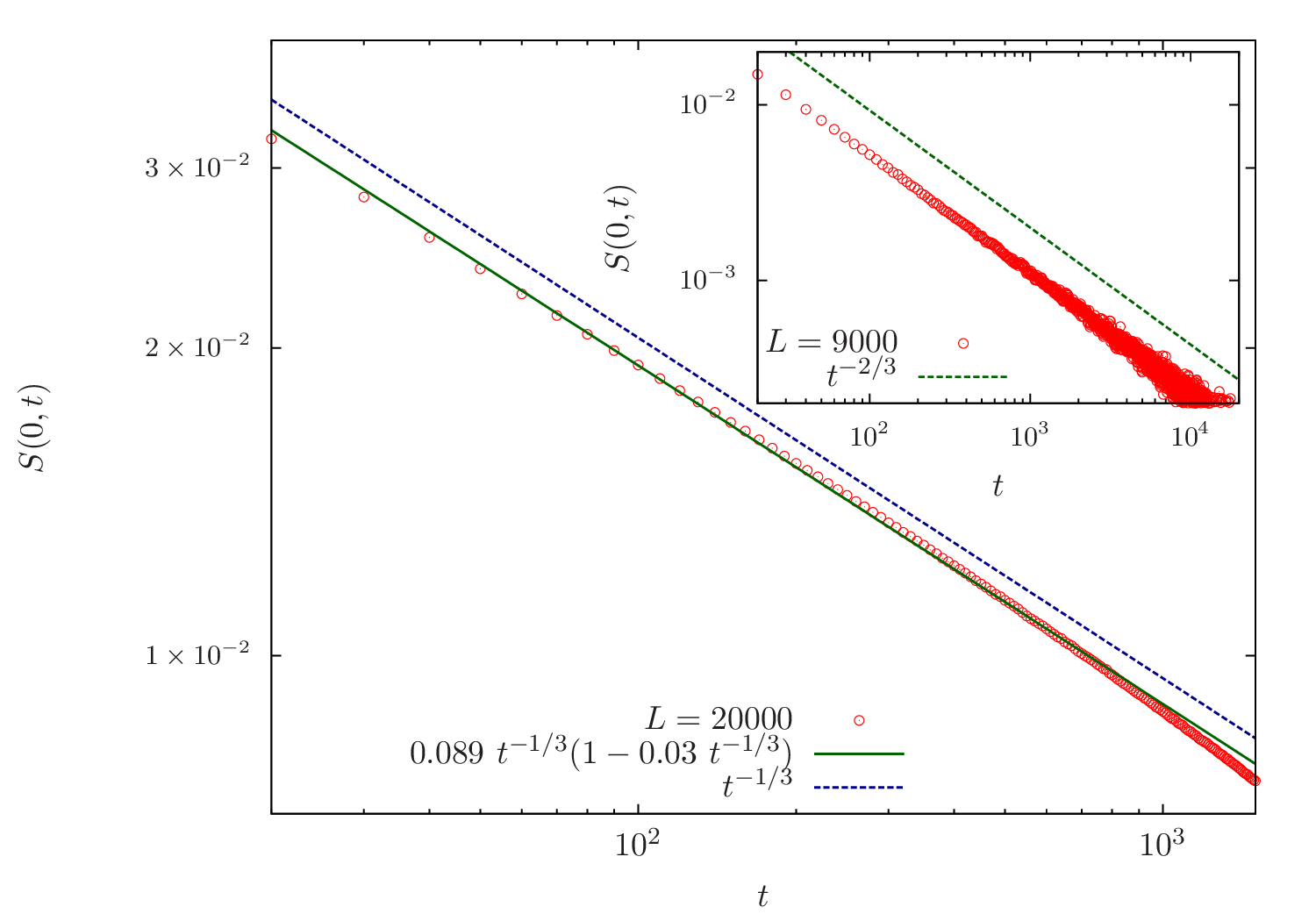}
\end{center}
\caption{Autocorrelation function at the critical point for the disordered exclusion process discussed in Sec.~\ref{sub_dis} for $\gamma=0.5$ (main) and $4$ (inset). The data are averaged over $2000$ independent disorder configurations for the rates chosen from (\ref{disdist}) with ${\tilde u}=1/2$.}
\label{fig_dis}
\end{figure}

Following similar steps to the models considered before, the stationary state current for a given configuration of particle hop rates is found to be $J= \rho \omega$, and the speed of the density fluctuations can be calculated using (\ref{disconser}) and its derivative with respect to $\omega$. In numerical simulations, for a given set of $u_i$'s, we calculated the speed and measured the autocorrelation function in (\ref{defnS2}). The numerical data were averaged over many disorder configurations to yield $\overline{S(0,t)}$ which is shown in Fig.~\ref{fig_dis}, and we find that our numerical results are consistent with (\ref{Srtscaling}) where the roughness exponent is as quoted above and the dynamic exponent $z=3/2$.

\section{CONCLUSIONS}
\label{concl}

Although the critical dynamics of some nonequilibrium phase transitions such as absorbing state transitions \cite{Hinrichsen:2000} have been rather well studied, such questions have not been addressed for the condensation transition and the related jamming transition barring some exceptions  \cite{Jain:2003,Godreche:2005}. The dynamical behavior in the stationary state for the latter class of transitions is the subject of this article.

Here we studied several one-dimensional models in which the particles have hard core interactions and always hop in one direction. Our main finding is that when the equal time density-density correlation function decays faster than the inverse of the distance, the autocorrelation function decreases with time in the same way as in the TASEP. However for slower decaying $S(r,0)$, the autocorrelation function $S(0,t)$ also decays slower than $t^{-2/3}$ and with an exponent that varies continuously with a model parameter. In a driven system of hard core particles with nonlocal hop rates, the static correlation function is found to decay as $r^{-2}$ and the dynamic exponent has been shown to be one \cite{Popkov:2011}, unlike here where $z=3/2$ is found to be robust. 

For a generalisation of the hop rate (\ref{hoprate}) given by  $u(n)=1+(b/n^\lambda)$, the jamming transition occurs when $0 < \lambda < 1, b > 0$ (besides the case considered here, namely, $\lambda=1, b > 2$) \cite{Armendariz:2013}. In the former case, at the critical point, the single site mass distribution $p(m) \propto \exp\left[- b (1-\lambda)^{-1} m^{1-\lambda} \right]$ which decays faster than a power law as in the fluid phase. As a consequence, the roughness exponent remains one half \cite{Priyanka:2014} and the dynamical behavior is the same as in the TASEP for all $0 < \lambda < 1, b > 0$.
We have also studied a bidirectional version of the model in Sec.~\ref{sub_model1} in which a particle may hop to either side with equal probability at a rate that depends on the number of holes in the direction toward which it chooses to hop. This model can be mapped to a symmetric ZRP whose steady state is the same as in Sec.~\ref{sub_model1}. As a result, the roughness exponent does not change from the driven model. However, since the steady state current vanishes, the dynamic exponent $z=2$ here \cite{Kriecherbauer:2010,Halpin-Healy:2015}. Our numerical simulations for the autocorrelation function  given by (\ref{Srtscaling}) are consistent with these values of the roughness exponent and the dynamic exponent.

We note that the Galilean invariance which holds for the KPZ equation and  yields the scaling relation $\alpha+z=2$ is violated for the models studied here at the critical point when the static correlation function decays slower than $1/r$. Precisely what is responsible for this breakdown is not understood.  We have been able to obtain an understanding of the KPZ dynamic exponent $z=3/2$ in the models discussed here using a hydrodynamic theory when the third cumulant of the hole distribution is finite, but an  extension of the hydrodynamic description to the complete range of parameters at the critical point is desirable. \\

{\bf Acknowledgements:} The authors thank M. Barma, J. Krug, D. Mukamel and G.M. Sch{\"u}tz for helpful discussions. Priyanka acknowledges the University Grants Commission for support through a research fellowship.

\clearpage

\appendix 


\section{Hydrodynamic equation for the height profile}
\label{app_hydro}

For the lattice-gas models considered here, the average particle density $\rho_i(t)=\langle \eta_i(t) \rangle$ at site $i$ obeys a continuity equation, ${\dot \rho_i(t)}=j_{i}(t)-j_{i-1}(t)$ where $j_i(t)$ is the average current in the bond connecting the sites $i$ and $i+1$. On a coarse-grained level, the local density (now defined in continuous space) obeys the following equation 
\be
\frac{\partial \rho(r,t)}{\partial t}+\frac{\partial J(\rho(r,t))}{\partial r}=0 ~,
\ee
where, assuming that local stationarity holds, the current depends on the space variable through the density and given by the expression in the stationary state \cite{Kriecherbauer:2010,Halpin-Healy:2015}. Writing $\rho(r,t)=\rho+ \delta \rho(r,t)$ and expanding the current to quadratic order in the deviation $\delta \rho(r,t)$ about the mean density, we obtain $J(\rho(r,t))=J(\rho)+ v \delta \rho(r,t) + (\lambda/2) (\delta \rho(r,t))^2 $ where
\bea
v &=& \frac{\partial J}{\partial \rho} \\
\lambda &=& \frac{\partial^2 J}{\partial \rho^2} ~.
\eea
Since the lowest order term in $\delta \rho$ obeys the following first order wave equation 
\be
\frac{\partial \delta \rho(r,t)}{\partial t}+v \frac{\partial \delta \rho(r,t)}{\partial r}=0 ~,
\ee
the density fluctuations move with a speed $v$.  
Using the mapping (\ref{hnmap}) and retaining the quadratic term that carries information about the decay of the density fluctuations, we obtain the KPZ equation for the height profile in one (space) dimension \cite{Kriecherbauer:2010,Halpin-Healy:2015}:
\be
\frac{\partial h}{\partial t}= -v\left(\frac{\partial h}{\partial r} \right)+ \frac{\lambda}{2} \left(\frac{\partial h}{\partial r} \right)^{2}+\nu \frac{\partial^2 h}{\partial r^2}+\zeta (r,t) ~,
\ee
where the noise is assumed to be white in both space and time with $\langle \zeta(r,t) \zeta(r',t') \rangle= D \delta(r-r') \delta(t-t')$ and $C(r,0)=(D/\nu) |r|$.

For the exclusion process described in Sec.~\ref{model}, we now consider the above equation for $\rho \geq \rho_c$. As explained in Sec.~\ref{unequal}, the speed $v$ given by (\ref{kinspeed}) is related to the variance of the mass. The coefficient $\lambda$ above is however related to the third cumulant of mass, $\kappa_3$:
\be
\lambda= \frac{\omega}{(\rho \sigma^2)^3} \left( \sigma^2 +2 \rho \sigma^4 - \kappa_3 \right)- \frac{2 \omega}{\rho^2 \sigma^2} ~.
\ee
For $\rho > \rho_c$, since the mass distribution is an exponential, all the mass cumulants exist and therefore, using the KPZ exponents, it follows that  the autocorrelation function decays as $t^{-2/3}$ in the fluid phase.

We now ask if the KPZ equation describes the dynamics at the critical point as well. As for the speed, we approach the critical point from the high density side to find the coefficient $\lambda$. 
In the vicinity of the critical point, for non-integer $b$, the generating function $g(\omega)$ for the weight factor $f(m)$ in (\ref{fmzrp}) is given by \cite{Priyanka:2014}
\be
g(\omega)= g(1)-s g'(1)+ \frac{s^2}{2!} g''(1)+...+\frac{(-s)^n}{n!} g^{(n)}(1)+ \alpha s^{b-1}+{\cal O}(s^{b})  ~,
\ee
where $s=1-\omega$, $n$ is the integer part of $b-1$ and $\alpha=b \pi \csc(b \pi)$. Using this in the density conservation equation (\ref{conser}), we have \cite{Priyanka:2014}
\be
{\frac{1}{\rho}=}
\begin{cases}
\frac{1}{\rho_c}- \frac{\alpha (b-1) g'(1)}{g^2(1)}~s^{b-2} +{\cal O}(s)&,~2 < b < 3 \\ 
\frac{1}{\rho_c}+ \frac{g'(1)}{g(1)} 
\left(\frac{g'(1)}{g(1)}-\frac{g''(1)}{g'(1)} -1\right) s - \frac{(b-1) \alpha}{g(1)} s^{b-2}& ,~3 < b < 4 \\
 \frac{1}{\rho_c}+ \frac{g'(1)}{g(1)} 
\left(\frac{g'(1)}{g(1)}-\frac{g''(1)}{g'(1)} -1\right) s+{\cal O}(s^2)& ,~b > 4 ~.
\end{cases}
\label{bg2}
\ee
On inverting this equation for the fugacity to lowest order in the deviation $\epsilon=\rho-\rho_c$, 
and using it in (\ref{currentexpr}) for the current, we get 
\be
{J(\rho) =} 
\begin{cases}
\rho_c+ v \epsilon+ {\cal O}(\epsilon^{\frac{1}{b-2}}) &,~ 2 < b < 3 \\
\rho_c+ v \epsilon+ {\cal O}(\epsilon^{b-2}) &,~ 3 < b < 4 \\
\rho_c+ v \epsilon+ {\cal O}(\epsilon^2)  &,~ b > 4 ~,
\end{cases}
\label{hyrdo2}
\ee
where $v$ is given by (\ref{kinspeed}). The above equation shows that the coefficient $\lambda$ is finite for $b > 4$ but diverges otherwise at the critical point.


\section{Current-density relation for the exclusion process with hole-dependent rates}
\label{app_curr}

For the current (\ref{currentexpr}), the speed $v=\omega+ \rho \partial \omega/\partial \rho$. Taking the derivative with respect to $\rho$ on both sides of (\ref{conser}) and using the identities for the Gauss hypergeometric function when $\omega=1$ \cite{Abramowitz:1964}, we  arrive at the speed $v_+$ at the critical point quoted in (\ref{kinspeed}). This speed changes sign when the expression for it for $b > 3$ vanishes {\it i.e.} $b^*=3 + \sqrt{2}$.

For $b < b^*$, since the slope $v_+(\rho_c)$ of the current at $\rho_c$ from the high density side is positive, the current continues to increase beyond the critical density. The density $\rho^* > \rho_c$ where the current is maximum can be found by setting the derivative of the current equal to zero and using the density-fugacity relation (\ref{conser}). For $b=2.5$ for which $\rho_c=1/3$, we find that $\rho^* \approx 0.65$ as shown in Fig.~\ref{fig_current}. For $b > b^*$, the current must decrease as the critical density is crossed since the speed $v_+$ is negative. But as the slope $v_{-}$ is positive, the maximum current is obtained at the critical density. Note that the derivative of the current does not exist at the critical point for $b > 3$.


\section{Steady state of the misanthrope process}
\label{app_misan}

For the hop rates given by (\ref{ummmisa}) and (\ref{vmmisa}), the steady state weight
factor $f(m)$ and its generating function $g(\omega)$ work out to be 
\bea
f(m) &=& \frac{c}{4} ~\frac{ C^{m-1} ((m+1)!)^2}{m! (c)_m} ~,~m \geq 1
\label{fmmis1} \\
g(\omega) &=& \frac{1}{2} \left[ 2-v(0)+ v(0) ~{_2}F_{1}(2,2,c;C \omega) \right] ~,
\label{gzmis1}
\eea
where  $(a)_m= a (a+1) ... (a+m-1)$ is the Pochammer symbol, $f(0)=1$, and $c$ and $C$ are given by 
\bea
c &=& \frac{3-2 v(0)}{1-v(0)} \\
C &=& \frac{c}{2 v(0)} ~.
\eea
As for the ZRP discussed in Sec.~\ref{sub_model1}, the critical density for the misanthrope process is found by setting $\omega=1/C$ in the 
density-fugacity relation (\ref{conser}), and we find that
\be
\varrho_c= \frac{4 v(0)}{c}~ \frac{{_2}F_{1}(3,3,c+1;1)}{2-v(0)+v(0)
  {_2}F_{1}(2,2,c;1)} ~.
\label{vrhmisa}
\ee
Note that the expression for the critical density in \cite{Evans:2014}
is incorrect since it is obtained by assuming that $c=4, C=1$ (see their (45)) which implies that $v(0)=2$ in contradiction with (\ref{vmmisa}).

The stationary state current in the lattice gas model with hop rates (\ref{ummmisa}) and (\ref{vmmisa}) can be written as
\bea
J(\rho)&=&\langle u(1,0) \eta_{i-2} (1-\eta_{i-1}) \eta_i \eta_{i+1} + u(2,0) \eta_{i-3} (1-\eta_{i-2}) (1-\eta_{i-1}) \eta_i \eta_{i+1} \nonumber\\
& & +...+u(1,1)\eta_{i-2}(1-\eta_{i-1})\eta_i (1-\eta_{i+1})\eta_{i+2}+...\rangle \nonumber \\
&=&\langle  \sum_{m=1}^{\infty}\sum_{m'=0}^{\infty} u(m,m') \eta_{i-(m+1)} \eta_{i} \eta_{i+(m'+1)}
\prod_{k=1}^{m}(1-\eta_{i-k}) \prod_{j=1}^{m'}(1-\eta_{i+j}) \rangle\\
&=& \rho \sum_{m=1}^{\infty} \sum_{m'=0}^{\infty} (v(m)-v(0)) v(m')  p(m) p(m') \\
&=& \rho \omega \frac{v(1)-v(0)}{v(0)}\left( \frac{2 v(0) _2F_1(1,3,c,C\omega)}{2-v(0)
+v(0) _2F_1(2,2,c,C\omega)}\right)^2 ~,
\eea
where we have used (\ref{pm1}) for the single site mass distribution and that the ratio 
\be
\frac{f(m)}{f(m-1)}= \frac{v(m-1)}{v(0)} ~\frac{v(1)-v(0)}{v(m)-v(0)}            ~, m \geq 1 ~.
\ee
The speed of the density fluctuations $v=\partial J/\partial \rho$ is given by a rather involved expression which we omit here, but it can be calculated easily using the above equation and the density-fugacity relation (\ref{conser}) numerically. At the critical point, since the single-site mass distribution given by 
\be
p(m)= \frac{v(0) (m+1) (m+1)!}{ \left[2- v(0)+v(0)~{_2}F_{1}(2,2,c;1) \right] ~(c)_m} ~,~m > 0
\ee
decays algebraically as $\sim m^{3-c}$, the mass fluctuations diverge in the thermodynamic limit when $5 < c< 6$ and remain finite otherwise. Thus, using (\ref{variance}), we find that the speed is discontinuous at the critical point for $c > 6$. 

\section{Static correlation function for the disordered exclusion process}
\label{app_dis}

When the hop rates depend on the particle, analogous to (\ref{homocorr}), we can write 
\bea
\langle \eta_i(0) \eta_{i+r}(0) \rangle = \sum_{\alpha=1}^N \textrm{Prob(site $i$
  is occupied by particle $\alpha$)} \times
\sum_{s=1}^{r} P_{s} (r-s; \{u_\alpha,...,u_{\alpha+s-1} \})  
\eea
where $P_{s} (r-s; \{u_\alpha,...,u_{\alpha+s-1} \})$ denotes the distribution of mass $r-s$ on $s$ consecutive sites in the ZRP when the sites have a hop rate $\{u_\alpha,...,u_{\alpha+s-1} \}$. Similar to (\ref{sumSr0}), this distribution is given by 
\be
P_{s} (r-s; \{u_\alpha,...,u_{\alpha+s-1} \})=\sum_{m_1,..,m_s} \prod_{j=\alpha}^{\alpha+s-1} \frac{\omega^{m_j} f_{j}(m_{j})}{ g_j(\omega)}~\delta_{\sum_{k=\alpha}^{\alpha+s-1} m_k,r-s} ~.
\ee
On averaging over the particle hop rates, we get 
\be
\overline{\langle \eta_i(0) \eta_{i+r}(0) \rangle}= \rho \sum_{s=1}^r \sum_{\{ m_j\}}
         \overline{p(m_1)} ~...~  \overline{p(m_s)} ~\delta_{\sum_{j=1}^s 
    m_j,r-s} ~,
\label{disordered}
\ee
where $\overline{p(m)}$ is the single site mass distribution obtained after
averaging over the disorder distribution and is given by
\bea
\overline{p(m)} = \int_{{\tilde u}}^1 du ~q(u) \frac{\omega^m f(m)}{g(\omega)} ~.
\eea
On inserting the  expression (\ref{disordered}) in  (\ref{homocorr}), we obtain (\ref{disGy}) in the main text.

\clearpage

\clearpage

\clearpage

\clearpage

\clearpage

\clearpage

\begin{figure}
\begin{center}
\includegraphics[width=1\textwidth]{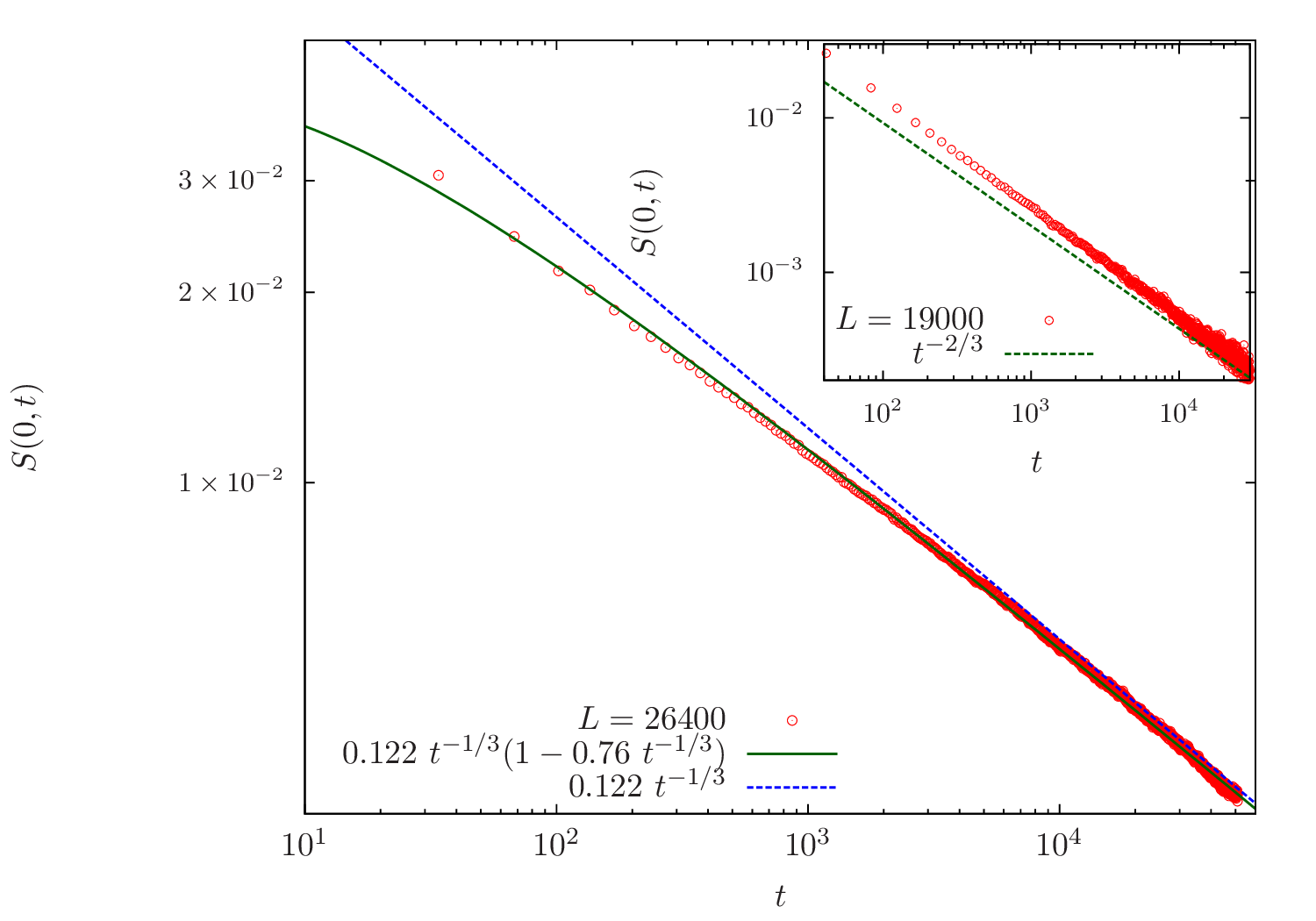}
\end{center}
\caption{Autocorrelation at the critical point for the generalised exclusion process discussed in Sec.~\ref{sub_misan} for  $c=5.5$ with $\rho_c=5/33, v=0.285714$ (main) and $c=9$ with $\rho_c=12/19, v=-0.00846192$ (inset).} 
\label{fig_misant}
\end{figure}

\clearpage


\clearpage



\end{document}